\documentclass[useAMS,usenatbib]{mn2e}
\usepackage{times,graphicx,float}
\usepackage{psfig}

\title[Modelling the `outliers' track of the radio--X-ray correlation]
  { Modelling the `outliers' track of the radio--X-ray correlation in X-ray binaries based on disc-corona model}
\author[C.-Y. Huang et al.]
  {Chang-Yin Huang$^{1,2}$, Qingwen Wu$^1$\thanks{Corresponding author, E-mail: qwwu@hust.edu.cn} and  Ding-Xiong Wang$^1$  \\
  $^1$School of Physics, Huazhong University of Science and Technology, Wuhan 430074, China\\
  $^2$College of Physical Science and Technology, Yangtze University, Jingzhou 434023, China\\
 }
\date{}

\pagerange{\pageref{firstpage}--\pageref{lastpage}} \pubyear{0000}

\def\LaTeX{L\kern-.36em\raise.3ex\hbox{a}\kern-.15em
    T\kern-.1667em\lower.7ex\hbox{E}\kern-.125emX}

\begin{document}

\label{firstpage}

\maketitle

\begin{abstract}
   The universal radio--X-ray correlation ($F_{\rm R}\propto F_{\rm X}^{b}$, $b\sim0.5-0.7$) has been found for a sample of black-hole X-ray binaries (BHBs) in their low/hard states, which can roughly be explained by the coupled model of jet and radiatively inefficient advection dominated accretion flow. However, more and more `outliers'  were found in last few years, which evidently deviate from the universal radio--X-ray correlation and usually show a much steeper correlation with an index of $\sim 1.4$. Based on simple physical assumptions, the radiatively efficient accretion flows are speculated to exist in these `outliers'. In this work, we test this issue by modelling the `outliers' track based on the radiatively efficient disc-corona model and the hybrid jet model. We find that our model predicts a steeper radio--X-ray correlation with slopes $\ga 1.2$ for the typical viscosity parameter of $\alpha\sim 0.05-0.2$. In particular, the slope is $\sim 1.4$ for the case of $\alpha\sim0.1$, which is consistent with the observational results of H1743$-$322 very well. Our results suggest that the `outliers' track may be regulated by the disc-corona model.

\end{abstract}


\section{Introduction}

  Most black-hole X-ray binaries (BHBs) are transient systems that undergo occasional outbursts, which
  display complex spectral and timing features. Normally, there are two main states in BHBs: high/soft (HS) state and low/hard (LH) state (e.g., \citealt{zdzi00,MR06}). The HS state is characterized by a strong thermal emission and a weak power-law component. However, the thermal emission is normally weak in the LH state and most of the radiation comes from the nonthermal power-law  component. Both the soft X-ray bumps observed in HS state of BHBs and optical/UV bumps observed in quasars can be naturally interpreted by the multi-temperature blackbody emission from a cold, optically thick, geometrically thin standard accretion disc (SSD; \citealt{SS73}). The prevalent accretion model for LH-state BHBs and low-luminosity active galactic nuclei (AGNs) is the hot, optically thin, geometrically thick advection-dominated accretion flows (ADAFs; also called radiatively inefficient accretion flows, RIAFs) that have been developed for black holes (BHs) accreting at low mass accretion rates (e.g., \citealt{ichi77,nara94,nara95,abra95}; see \citealt{kato08} and \citealt{nara08} for recent reviews).

  There is a strong connection between the radio and X-ray emission in the LH state of BHBs. The quasi-simultaneous radio and X-ray fluxes roughly follow a universal non-linear correlation \citep[$F_{\rm R}\propto F_{\rm X}^{b}$, $b\sim0.5-0.7$,][]{hann98,corb03,gall03,corb13}. This non-linear correlation seems to be maintained down to the quiescent state \citep{corb03,gall06,corb08}. By taking into account the BH mass, the relation was extended to AGNs, which is called ``fundamental plane'' of BH activity \citep{merl03,falc04,wangy06,kord06,li08,yuan09,gult09,plot12}. However, in recent years, several BHBs were found to lie well outside the scatter of the original radio--X-ray correlation (e.g., H1743$-$322, \citealt{jonk10,cori11}; Swift 1753.5$-$0127,  \citealt{cado07,sole10}; XTE J1650$-$500, \citealt{corb04}; XTE J1752$-$223, \citealt{ratt12}). These outliers roughly form a different track (`outliers' track) which follow a steeper correlation with an index of $b\sim 1.4$ as initially found in H1743$-$322 \citep{cori11}. Some of these sources (e.g., H1743$-$322, XTE J1752$-$223, MAXI J1659$-$152) jump to the standard universal correlation when they fade towards quiescence \citep{jonk10,jonk12,cori11,ratt12}.

  Although the jets have been observed in different kinds of high-energy objects (e.g., AGNs, BHBs, GRBs, etc.), the detailed physical mechanism for the jet formation is still unclear. The popular mechanisms for the jet production include Blandford-Znajek (BZ) process \citep{bz77} and Blandford-Payne (BP) process \citep{bp82}. Both mechanisms involve energy extraction via open large-scale magnetic fields either from a rotating BH in form of Poynting flux (BZ process) or from the rotating accretion disc in form of magnetically driven wind (BP process). The hybrid model proposed by \cite{meie99}, as a variant of BZ model, combines the BZ and BP effects through the large-scale magnetic fields threading the accretion disc outside the ergosphere and the rotating plasma within the ergosphere. This model seems to be supported by magnetohydrodynamic (MHD) simulations (e.g., \citealt{koid00,mcki04,hiro04,hawl06}) and some recent observations (e.g., \citealt{nemm07,wu08,wu11,lisn12}).

  The radiatively inefficient ADAF is expected to produce the X-ray emission with $L_{X}\propto \dot{M}^{q}$ ($\dot{M}$ is accretion rate, $q\sim 2.0$, e.g., \citealt{nara97,merl03,yuan05b,wu06}), which can roughly explain the universal radio--X-ray correlation ($L_{R}\propto L_{X}^{0.7}$) if considering the scaling between the jet luminosity and jet power, $L_{R}\propto Q_{\rm jet}^{1.4}$ (e.g., \citealt{bk79,falc96,hein03}) and $Q_{\rm jet}\propto \dot{M}$ (e.g., \citealt{falc95,wu11}). The detailed calculations based on the ADAF-jet model do support this scenario (e.g., \citealt{yuan05b}). From the simple physical assumptions, the `outliers' of BHBs can be understood if they are accreting through radiatively efficient accretion disc, where $L_{X}$ is roughly proportional to $\dot{M}$. The standard disc-corona model \citep{SS73,liu02a} and the luminous hot accretion flow \citep{yuan01} are two possible candidates for the radiatively efficient accretion flow. Recently, there are some observational evidences point to the possible presence of a cool inner disc in the bright hard state of BHBs if its luminosity is higher than $\sim 0.1\%$ of Eddington luminosity (e.g., \citealt{mill06a,mill06b,ryko07,rama07,reis09,reis10}, but see also \citealt{done10,pla13} for different opinion). From the theoretical point, \cite{liu07} also found that there may exist an inner cool disc in the LH state of BHBs due to the condensation of the matter from the ADAF if its accretion rate is larger than 0.1\% of Eddington rate.

  The `outliers' are normally in bright hard state, and the steeper radio--X-ray correlation that regulated by the radiatively efficient model has been discussed in former works (e.g., \citealt{merl03,falc04,cori11,corb13}). In this work, we present detailed calculations based on a jet that formed from the disc-corona system, and test whether this model can explain the radio--X-ray correlation of the `outliers' track or not. In Section 2, we present the disc-corona model and the jet model. The theoretical result and its comparison with observations are presented in Section 3. We discuss our results in Section 4.

\section{The model}

   \subsection{Disc-corona model}

  According to typical disc-corona model, part of the viscously dissipated energy, $Q_{\rm dissi}^{+}$, is released in the disc, $Q_{\rm d}^{+}$, and emit eventually as blackbody radiation. The rest dissipated energy, $Q_{\rm c}^{+}$, is transported into the corona by the magnetic field. The gravitational power dissipated in unit surface area of the thin accretion disc surrounding a Kerr black hole is
  \begin{equation}\label{e1}
  Q_{\rm dissi}^{+}=\frac{3GM_{\rm BH}\dot{M}}{8\pi R^3}\frac{\mathcal{L}}{\mathcal{B}\mathcal{C}^{1/2}},
  \end{equation}
  where $M_{\rm BH}$ is BH mass, $\mathcal{B}$, $\mathcal{C}$, and $\mathcal{L}$ are the general relativistic correction factors for the standard disc \citep{nt73,pt74}. The energy equation for the cold disc is
  \begin{equation}\label{e2}
  Q_{\rm d}^{+}=Q_{\rm dissi}^{+}-Q_{\rm c}^{+}=\frac{4\sigma T_{\rm d}^{4}}{3\tau_{\rm d}},
  \end{equation}
  where $T_{\rm d}$ is effective temperature in the mid-plane of the disc and $\tau_{\rm d}=\tau_{\rm es}+\tau_{\rm ff}$ is optical depth in the vertical direction of the disc.

The corona is assumed to be heated by the reconnection of the magnetic fields that generated by the buoyancy instability in the disc (e.g., \citealt{dima98,liu02a}). The power dissipated in the corona is
 \begin{equation}\label{e3}
    Q_{\rm c}^{+}=\frac{B_{\rm d}^2}{4\pi}V_{\rm A},
  \end{equation}
  where $B_{\rm d}$ is the tangled small-scale magnetic field in the disc and $V_{\rm A}=B_{\rm d}/\sqrt{4\pi\rho}$ is the Alfv$\acute{\rm e}$n speed ($\rho$ is mass density of the disc). As the detailed physics for generating magnetic fields in the accretion disc are still quite unclear, the so-called `$\alpha$-prescription' is widely adopted in most of the works on accretion discs \citep{SS73}, of which the magnetic stress tensor $t_{r\varphi}$ is assumed to be proportional to the total pressure ($P_{\rm tot} = P_{\rm gas} + P_{\rm rad}$), or gas pressure $P_{\rm gas}$ or $\sqrt{P_{\rm gas}P_{\rm tot}}$ \citep{saki81,stel84,taam84}. The third prescription seems to be supported by stability analysis and recent observations (e.g., \citealt{blae01,cao09}). Therefore, we adopt this magnetic stress tensor in our calculation
  \begin{equation}\label{e4}
    t_{r\varphi}=P_{\rm mag}=\frac{B_{\rm d}^2}{8\pi}=\alpha \sqrt{P_{\rm gas}P_{\rm tot}},
   \end{equation}
  where $P_{\rm mag}$ is magnetic pressure in the disc and $\alpha$ is the viscosity parameter of the cold disc.

  The continuity equation of the disc is
  \begin{equation}
   \dot{M}=-4\pi R H_{\rm d}\rho V_{\rm r} \mathcal{D}^{1/2},
 \end{equation}
 where  $H_{\rm d}$ is half thickness of the disc, $V_{\rm r}$ is radial velocity of the accretion flow at radius $R$, and $\mathcal{D}$ is general relativistic correction factors \citep{nt73}.

  The angular-momentum equation for the disc is \citep{nt73}
 \begin{equation}
 4\pi H_{\rm d} t_{r\varphi}= \dot{M}\sqrt{\frac{GM_{\rm BH}}{R^3}}\frac{\mathcal{F}}{\mathcal{D}},
 \end{equation}
 where the factor $\mathcal{F}$ is a function of $\mathcal{D}$ as introduced in \cite{riff95}.

   The equation of state for the gas in the disc is
   \begin{equation}
    P_{\rm tot}=P_{\rm rad}+P_{\rm gas}=\frac{1}{3}a_0 T_{\rm d}^4+\frac{\rho k T_{\rm d}}{\mu m_{\rm p}},
    \end{equation}
    where $a_0$, $k$, $m_{\rm p}$ and $\mu$ are respectively radiation constant, Boltzman constant, proton mass and mean atomic mass ($\mu=0.62$ is adopted).

  The energy released into the corona, $Q_{\rm c}^{+}$, is balanced via the inverse Compton scattering by the soft photons from the under thin disc. Thus we have \citep{liu02a}
  \begin{equation}\label{e5}
    \frac{B_{\rm d}^2}{4\pi}V_{\rm A}=\frac{4kT_{\rm e}}{m_{\rm e}} \tau U_{\rm rad},
  \end{equation}
   where $U_{\rm rad}=a_0 T_{\rm d}^4$, $\tau$, $T_{\rm e}$ and $m_{\rm e}$ are energy density of the soft photon field, optical depth of the corona, electron temperature and electron mass respectively.

  Solving equations (1)--(7) numerically, we can obtain self-consistently global solutions of the cold disc. For a given optical depth $\tau$, the electron temperature of the corona at given radius can be derived from equation (8). Spectral studies of BHBs in the LH state indicate that the value of $\tau$ is around $\la$0.5--2 (e.g., \citealt{gier97,zdzi99,tori11}). Theoretical studies of the disc-corona model also reveal that the value of $\tau$ lies in the range of 0.1--0.8 (e.g., \citealt{liu02a,liu03,liu12,yao05,cao09,qiao12,you12}). In this work, we consider a slab corona with the optical depth $\tau\sim0.1-1$, the height $H_{\rm c}=20R_{\rm g}$ \citep{liu02a}, the inner boundary $R_{\rm in}=R_{\rm ms}$ and the outer boundary $R_{\rm out}=100R_{\rm g}$, where $R_{\rm g}\equiv GM_{\rm BH}/c^2$ and $R_{\rm ms}$ are gravitational radius and radius of the innermost stable circular orbit (ISCO), respectively.

   The radiative spectrum of the disc-corona system was simulated by using Monte Carlo method that developed in our former work \citep[see][for more details]{gan09}. We briefly summarized it here, which mainly includes four steps: i) sample a seed photon from the cold disc based on the Plankian spectrum for a given $T_{\rm d}$ at $R$; ii) draw a value for its free path and test whether it can leave the corona; iii) simulate the interaction of the photon with the electrons in the corona; iv) repeat steps (ii) and (iii) till the photon leaves the system of the corona. The X-ray luminosity can be derived by integrating the spectrum.

   \subsection{Jet model}

   The jet power, $Q_{\rm jet}$, is estimated based on the hybrid jet model proposed by \cite{meie99,meie01},
   \label{sect:Obs}
    \begin{equation}\label{eq3}
    Q_{\rm jet}=B_{\rm p}^2 R_{\rm J}^4 \Omega^2/32c,
   \end{equation}
   where $R_{\rm J}$ is characteristic size of the jet formation region, and $B_{\rm p}$ and $\Omega$ are respectively the poloidal magnetic field strength and angular velocity of the disc in that region. In this work, we assume that the poloidal field $B_{\rm p}$ is a fraction of the total magnetic field $B_{\rm d}$ of the disc, e.g., $B_{\rm p}=fB_{\rm d}$ with $0<f<1$ (\citealt{livi99,livi03}).  \cite{livi03} proposed the  poloidal magnetic field $B_{\rm p}\sim B_{\rm d} (H/R)^{1/2}$ by modeling the variability of GRS 1915+105, which roughly corresponds to the parameter $f\sim0.1-0.2$ for the standard thin disc. Some numerical simulations also indicate that the strength of poloidal magnetic field threading the disc is a small fraction of the total field (e.g., $f\sim0.2$ in \citealt{hawl96,hawl11,ghos97,hawl00}). Therefore, we adopt $f=0.2$ in our calculations. Following \cite{meie01} and \cite{nemm07}, we adopt the size of the jet formation region $R_{\rm J}$ equal to the radius where the magnetic field strength and radiation flux of the disc reach the maximum ($\sim 1.5 R_{\rm ms}$), and all quantities are evaluated at this radius.

   In order to compare with the observational results directly, we need to convert the theoretical jet power to radio luminosity or convert the observed radio luminosity to jet power. According to  optically thick jet model (e.g., \citealt{bk79,falc96,hein03}), the radio luminosity $L_{\rm R}$ is related to the jet power in a form of $L_{\rm R}\propto Q_{\rm jet}^{17/12}$. Based on the estimation of jet power for Cyg X-1 and GRS 1915+105, \cite{hein05} suggested that the normalization is $\sim6.1\times10^{-23}$, i.e.,
   \label{sect:Obs}
   \begin{equation}\label{eq5}
    L_{\rm R}=6.1\times10^{-23} Q_{\rm jet}^{17/12} \ \mathrm{erg \ s^{-1}}.
    \end{equation}
    We adopt this relation in converting the jet power to radio luminosity of the jet.

   \begin{figure}\label{fig1}
   \centering
   \includegraphics[width=8cm]{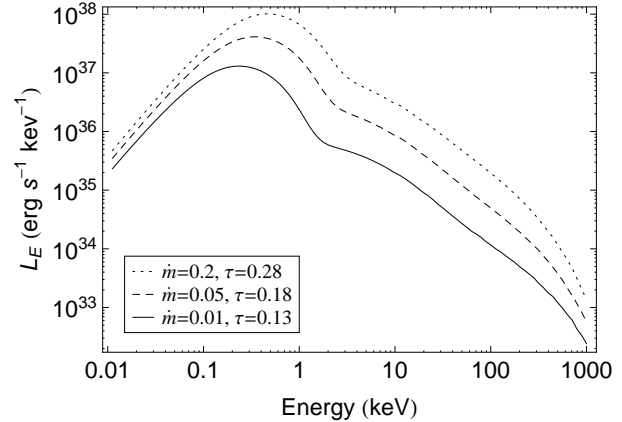}

   \caption{The typical hard spectra with photon index $\Gamma\simeq1.8$ reproduced by our disc-corona model, where $M_{\rm BH}=10M_{\odot}$, $a_*=0.5$ and $\alpha=0.1$ are adopted, and $\dot{m}\equiv\dot{M}/\dot{M}_{\rm Edd}$ is the accretion rate in unit of Eddington rate. }
  \end{figure}

\section{Results}

  Since the photon index of the X-ray spectrum is determined by both the coronal electron temperature and optical depth, we can reproduce the hard X-ray spectrum as that observed in LH-state ¡®outliers¡¯ by adjusting the proper optical depth of the corona for a given accretion rate. For example, we present the spectra with the hard X-ray photon index $\Gamma\simeq1.8$ for our disc-corona model in Figure 1, where different optical depths are chosen for different accretion rates in the cases of $a_*=0.5$ and $M_{\rm BH}=10M_\odot$.  We note that the 1--10 keV X-ray luminosity of our model is not very sensitive to the value of optical depth (e.g., $0.1<\tau<1$) when the dissipated energy in the corona is fixed (see equation 3), even there are some changes in the shape of the spectrum. For simplicity, we fix $\tau=0.5$ in exploring the radio--X-ray correlations.

   The radio--X-ray relations for different viscosity parameters of $\alpha=0.05,0.1$ and 0.2 are presented in the top panel of Figure 2 for given typical BH mass of $10M_\odot$ and BH spin of $a_*=0.5$. It can be found that the radio and X-ray luminosities are positively correlated even though they do not follow a simple power-law correlation, where the relation become a little bit steeper at high luminosities for given $\alpha$ parameter. Furthermore, a larger $\alpha$ value leads to a steeper radio--X-ray correlation. We find that the slope of the radio--X-ray relation is larger than 1.2 even for the case of $\alpha=0.05$. For comparison, the thick solid line with the slope of 1.4 is plotted in top panel of Figure 2, which is roughly consistent with the case of $\alpha=0.1$ for the typical luminosity range of the `outliers' track with $4\times 10^{36}\rm erg \ s^{-1}\la\it L_{\rm X}\la\rm 10^{38}\rm erg \ s^{-1}$. In the bottom panel of Figure 2, we show the radio--X-ray correlation for different BH spins at the given viscosity parameter $\alpha=0.1$. The larger BH spin parameter leads to the higher radio luminosity (or jet power), since that the jet power is more sensitive to the BH spin parameter compared with the X-ray emission (or disc luminosity).

   In Figure 3, we compare our theoretical results with the observed radio--X-ray correlation for the `outliers' track, where all the observational data are taken from \citet[and references therein]{corb13} and the sources that stay in the universal correlation are also included. We find that the observed radio--X-ray correlation of `outliers' track can be roughly reproduced with $\alpha\sim0.1$, which is not very sensitive to BH spin parameters. For comparison, we also plot a case of $\tau=0.3$ and $a_*=0.8$ (dotted line), which is roughly the same as that of a slightly larger optical depth $\tau=0.5$ with the same BH spin (thick-solid line). It suggests that our main conclusion will not change even if the different optical depth of the corona may exist for different luminosities. GRS 1915+105 slightly deviates from our above model predictions. However, it will be roughly consistent with model predictions if $M_{\rm BH}=15M_\odot$ is adopted (thin-solid line in Figure 3).

  \begin{figure}\label{fig1}
   \centering
   \includegraphics[width=8cm]{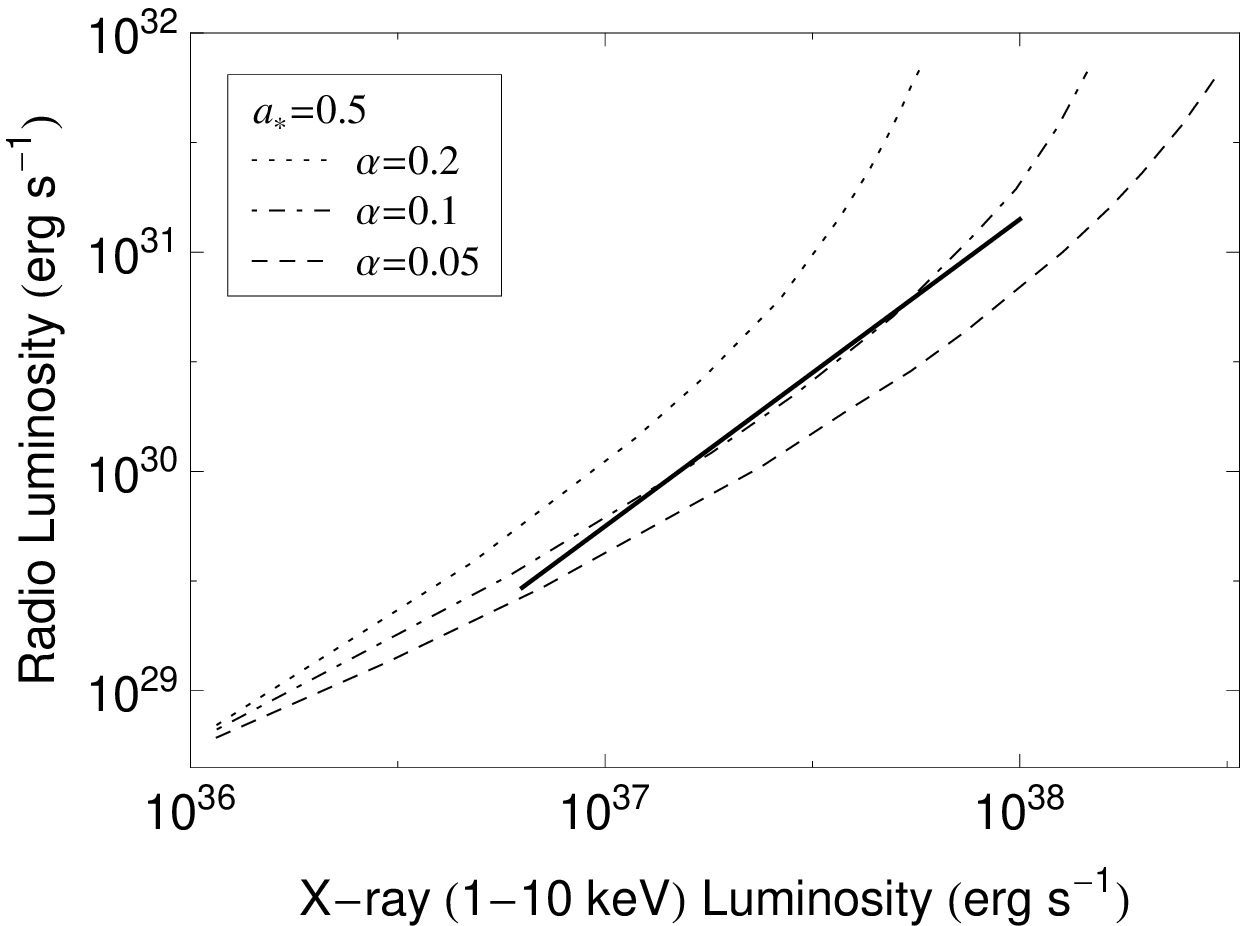}
   \vspace{0.5 cm}
   \includegraphics[width=8cm]{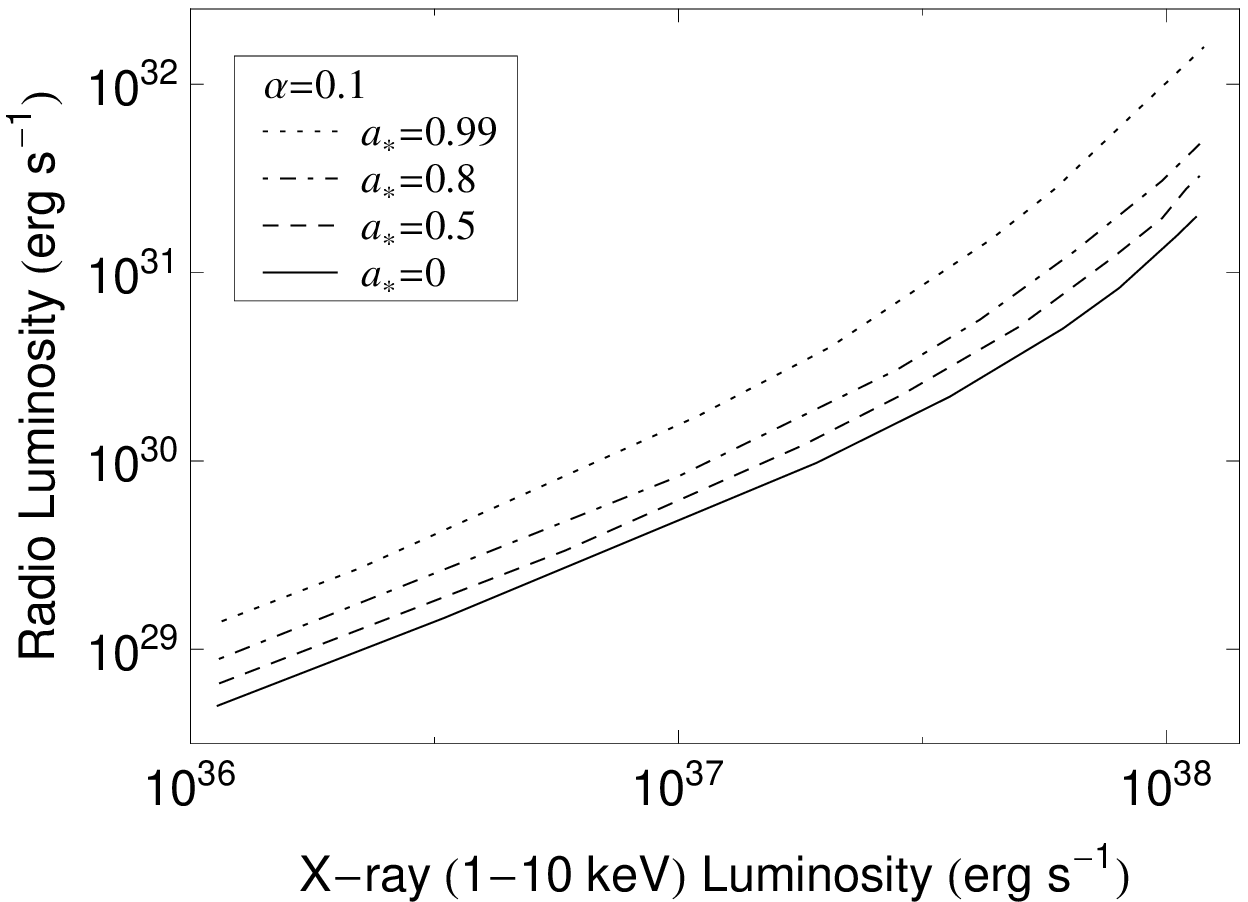}

   \caption{The top panel shows the radio--X-ray correlations based on our jet and disc-corona model for $\alpha=0.05$, 0.1, and 0.2 respectively at given typical BH mass of $10M_\odot$ and  BH spin $a_*=0.5$. For comparison, the straight line with slope of 1.4 is also plotted, which is shown by the thick solid line (top panel). The bottom panel shows the radio--X-ray correlations for different BH spins from $a_*=0$ to 0.99 for given $\alpha=0.1$.}
  \end{figure}

  \begin{figure*}
   \centering
   \includegraphics[width=15cm]{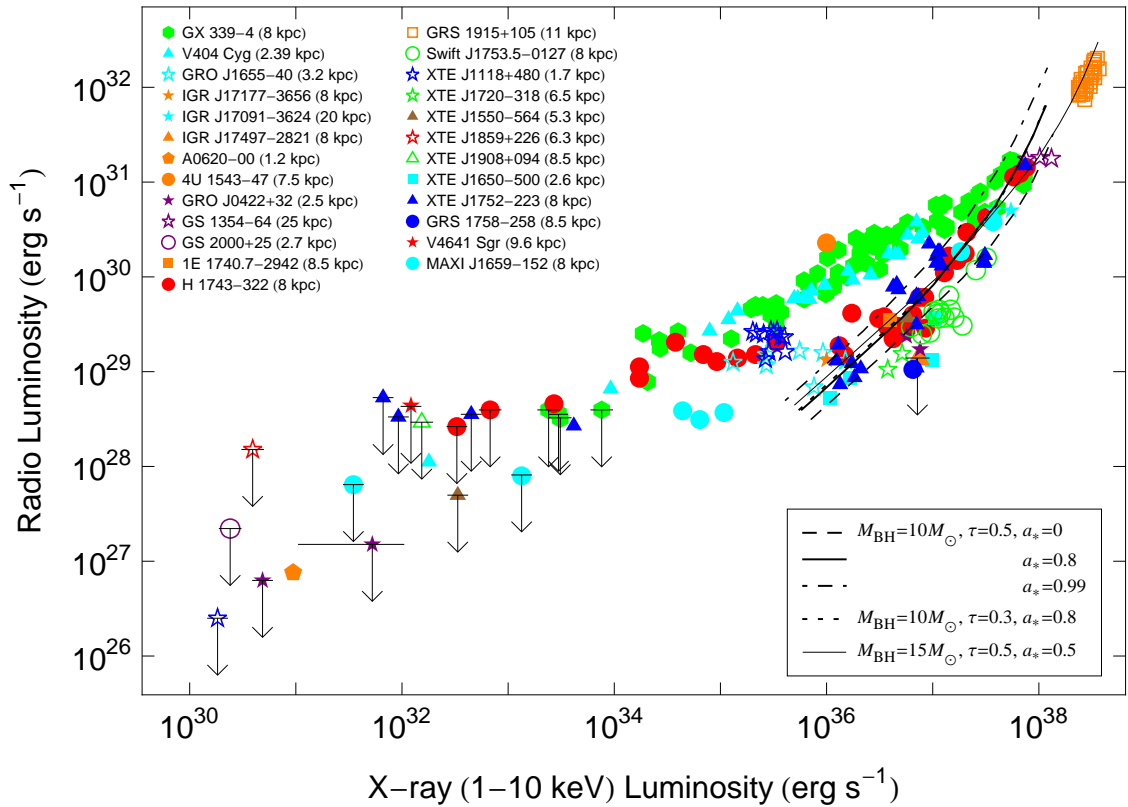}
   \caption{Comparison of the theoretical radio--X-ray correlations with that of observational results. The symbols with different colors denote the data from 25 hard-state BHBs which are taken from Corbel et al. (2013). The thick-solid, dashed and dot-dashed lines represent the correlations estimated by jet-disc/corona model for different BH spins at given $\alpha=0.1$, $\tau=0.5$ and $M_{\rm BH}=10M_\odot$. The correlation for a different optical depth of $\tau=0.3$ is also plotted by the dotted line which almost coincides with that of $\tau=0.5$ for the same BH spin. The thin-solid line denotes the correlation for $M_{\rm BH}=15M_\odot$, which may apply to the higher BH mass of GRS 1915+105.}
  \end{figure*}

\section{discussion}

   The universal radio--X-ray correlation of $F_{\rm R}\propto F_{\rm X}^{0.5-0.7}$ of BHBs in their LH states \citep[e.g.,][]{hann98,corb03,gall03,corb13} has been quantitatively explained
   by the coupled ADAF-jet model in \cite{yuan05b}, where the radio and X-ray emissions are dominated by the radiation from the jet and ADAF respectively. However, the physical mechanism for the `outliers' track is still unclear. In this work, we explore the radio--X-ray correlation based on the jet and disc-corona model, and find that this model can roughly reproduce the steeper correlation as found in `outliers' if assuming the jet launching and radiation behave identically in both tracks.

  \cite{wugu08} found that there exist two kinds of hard-state sources, where the hard X-ray photon index is anti-correlated to the luminosity for BHBs in the dim hard state, and, however, they become positively correlated for the sources in the bright hard state. The anti-correlation and positive correlation are consistent with the predictions of ADAF model and disc-corona model respectively (e.g., \citealt{cao07,you12,qiao13}). The critical Eddington ratio for the positive and anti-correlations is $\sim$ 1\%, which is also roughly consistent with the prediction for the disk transition \citep[][]{wugu08}. The hard X-ray photon index is also positively correlated with the luminosity for these `outliers', which is consistent with the prediction of disc-corona model \citep[][]{cw14}. Furthermore, the broad Fe K emission lines observed in the bright hard state of some `outliers' (e.g., H1743$-$322, \citealt{mccl09}; XTE J1650$-$500, \citealt{ross05}; XTE J1550$-$564, \citealt{sobc00}) suggest that the line emission region extends very close to ISCO and an inner cool disc may exist. GX 339-4 also has a positive correlation between the hard X-ray photon index and luminosity in its bright hard state \citep[][]{wugu08,cw14}, even it was suggested to form the universal correlation \citep[][]{cori11}. It is interesting to note that the radio--X-ray correlation also becomes steeper for GX 339--4 when the X-ray luminosity $L_{X}\ga 10^{37}\rm erg\ s^{-1}$ ($F_{\rm 8 GHz}\propto F_{\rm 3-9keV}^{\sim1.14}$, e.g., Figures 8 and 9 in \citealt{corb13} or Figure 1 in \citealt{cw14}). Therefore, all BHBs in bright hard state may follow the so-called `outliers' track.

  We find that the slopes of radio--X-ray correlation that based on the radiatively efficient disc-corona model are much steeper than the traditional slope of 0.7 even though it is not a simple power-law relation. For example, the slope is $\ga$ 1.2 even for the smallest adopted viscosity parameter $\alpha=0.05$. For a larger $\alpha$ value, the radio--X-ray correlation becomes steeper (see top panel of Figure 2), which is because the magnetic field (or jet power) becomes stronger (see Equation 4), while the X-ray emission (disc/corona) is not very sensitive to viscosity parameter. Even for a given $\alpha$ parameter, the radio--X-ray correlation also becomes steeper at high luminosities. The physical reason is that the radio emission (or jet power) is roughly proportional to the accretion rate while the corona (X-ray emission) roughly saturates at high accretion rate due to the stronger cooling (e.g., \citealt{liu09} Figure 1). We find that our model prediction with $\alpha\sim0.1$ can roughly reproduce the observed radio--X-ray correlation for this `outliers' track, where the $\alpha\sim0.1$ is a typical viscosity parameter that constrained from magnetohydrodynamic simulations (e.g., $\sim0.05-0.2$, \citealt{hawl02}).

   The radio--X-ray correlation of our model is not very sensitive to the BH spin parameter, because both the X-ray emission from the disc and the radio emission from the jet are related to BH spin.  We cannot constrain the BH spin for a given BHB through this comparison due to the uncertainties of our model and the uncertainties in converting the jet power to radio luminosity. Fortunately, the BH spin parameter does not affect the slope of the radio--X-ray correlation, which will not change our main conclusion (see bottom panel of Figure 2).

  In our disc-corona model, we find that the hard 1--10 keV X-ray emission is roughly proportional to $\dot{M}^{0.8}$. It suggests that the corona becomes weak with increasing $\dot{M}$ (e.g, $L_{\rm cor}/L_{\rm bol}\propto\dot{M}^{-0.2}$), which is qualitatively consistent with observations (see also \citealt{cao09,wangj04}). The magnetic field of the disc obeys the relation $B_{\rm d} \propto P_{\rm mag}^{0.5}\propto \dot{M}^{0.4}$ for the case of the adopted magnetic stress tensor. Considering the relation between jet power and radio luminosity, $L_{R}\propto Q_{\rm jet}^{17/12}$, for the theoretical jet model, we find $L_{R}\propto L_{\rm X, 1-10keV}^{1.4}$ for our model since the jet power $Q_{\rm jet} \propto B_{\rm d}^{2} \propto \dot{M}^{0.8} \propto L_{\rm X, 1-10keV}$. We note that the exact slope of radio--X-ray correlation is also affected by the viscosity parameter (see top panel of Figure 2). However, the slopes of the radio--X-ray correlation based on the jet-disc/corona model are generally larger than 1.2 for the typical viscosity parameter of 0.05--0.2, which are much steeper than those predicted by the ADAF-jet model (e.g., $\sim 0.7$, \citealt{yuan05b}). Therefore, our results with $\alpha\sim 0.1$ give a natural explanation for the radio--X-ray correlation of the `outliers', and support that this `outliers' track may be regulated by the disc-corona model as pointed out by \cite{cori11}.

   The radio--X-ray correlation based on the disc-corona model and ADAF model can roughly explain the `outliers' track and universal track respectively. The transition track may be regulated by a transition state of accretion flow. The radiative efficiency will increase very fast at a critical accretion rate, where the low radiative efficiency of ADAF will transit to high radiative efficiency of disk-corona. If this is the case, we expect the X-ray luminosity will increase much faster than the radio luminosity if the accretion rate approaches the critical rate for disk transition (e.g., $\sim$ 0.01 Eddington accretion rate), which can roughly explain the flat transition track (see Figure 3). It should be noted that different values of $\alpha$ will lead to different critical accretion rates for disc transition (e.g., \citealt{liu09,qiao09}). For a larger $\alpha$, the critical accretion rate will be large. This may be the reason why the universal radio--X-ray correlation can extend to higher X-ray luminosities in some BHBs (e.g., GX 339$-$4).

   \cite{yuan01} has shown that a hot accretion flow may exist when the accretion rate is larger than a critical rate, where this hot accretion flow may be also radiatively efficient (e.g., \citealt{xie12}). If this is the case, the jet launched from the hot accretion flow may also provide a possible explanation for the `outliers' track of the radio--X-ray correlation. More direct calculations are expected to test this scenario, which is beyond our work. However, it is still unclear whether the hot accretion flow with very high accretion rate exists or not if considering the possible condensation of the matter from the hot plasma to a cool disc, since \cite{liu07} found that the presence of a small cool disc is possible for the case of luminosity larger than 0.1\% of Eddington luminosity. Moreover, the broad Fe K emission lines observed in the bright hard state of some `outliers'do imply the existence of the inner cool disc, which cannot be explained by the hot accretion flows. We suggest that the high-sensitivity X-ray observations (e.g., XMM-Newton) on the cool disc for these `outliers' may help to test this issue.

\section*{Acknowledgments}

We thank the anonymous referee for constructive comments which greatly improved the manuscript. Q. Wu thanks Henk Spruit and Max-Planck Institute for Astrophysics for their hospitality where part of the work was done. This work is supported by the National Basic Research Program of China (2009CB824800), New Century Excellent Talents in University (NCET-13-0238) and the NSFC (grants 11103003, 11133005 and 11173011).

\label{lastpage}

\end{document}